\documentclass[twocolumn]{aastex631}

\usepackage{longtable}
\usepackage{float}
\usepackage{mathrsfs}
\usepackage{amsmath}
\usepackage{gensymb}

\newcommand{\macs}{MACS~J0717.5$+$3745}
\newcommand{\xmm}{XMM-{\it Newton}}
\newcommand{\chandra}{{\it Chandra}}
\newcommand{\TChan}{T$_{\text{Chandra}}$}
\newcommand{\TXMM}{T$_{\text{XMM}}$}
\newcommand{\TSZ}{$T_{\rm SZe}$}
\newcommand{\TSZm}{$T_{\rm rSZe}$}
\newcommand{\spire}{{\it Herschel}-SPIRE}
\newcommand{\spirefts}{SPIRE-FTS}
\newcommand{\herschel}{{\it Herschel}}
\newcommand{\planck}{{\it Planck}}
\newcommand{\Tdust}{T_{dust}}
\newcommand{\Tcmb}{T_{CMB}}
\newcommand{\RSZe}{rSZe}
\newcommand{\TSZe}{tSZe}
\newcommand{\KSZe}{kSZe}
\newcommand{\SZe}{SZe}
\newcommand{\msun}{M$_{\odot}$}
\newcommand{\szpack}{\texttt{SZPack}}
\newcommand{\RAhour}{$\overset{\text{h}}{.}\,$}
\newcommand{\RAmin}{$\overset{\text{m}}{.}\,$}
\newcommand{\RAsec}{$\overset{\text{s}}{.}\,$}
\newcommand{\declsec}{\arcsec\hspace{-4pt}.\hspace{2pt}}
\newcommand{\vpec}{$v_{\text{pec}}$}
\newcommand{\vpecm}{v_{\text{pec}}}
\newcommand{\ofts}{$O_{FTS}$}
\newcommand{\oftsm}{O_{FTS}}
\newcommand{\mjysr}{$\text{MJySr}^{-1}$}

\newcommand{\sigrsze}{$\sigma_{\rm rSZe}$}
\newcommand{\sigchan}{$\sigma_{\rm Chandra}$}
\newcommand{\sigxmm}{$\sigma_{\rm XMM}$}

\begin{document}

\title{Measuring the Temperature of Extremely Hot Shock-Heated Gas in the Major Merger MACS~J0717.5$+$3745 with Relativistic Corrections to the Sunyaev-Zel'dovich Effect}

\correspondingauthor{B.~Vaughan, {\tt bjv37@cornell.edu}}

\author[0000-0002-9813-0270]{Benjamin J. Vaughan}
\affiliation{Rochester Institute of Technology, \
1 Lomb Memorial Drive, \
Rochester, NY 14623, USA}
\affiliation{Cornell University, \
109 Clark Hall, \
Ithaca, NY 14850, USA}
\author[0000-0002-8213-3784]{Jack Sayers}
\affiliation{California Institute of Technology, \
1200 E California Blvd, \
Pasadena, CA 91125, USA}
\author[0000-0002-9941-2077]{Locke Spencer}
\affiliation{University of Lethbridge,
\ 4401 University Dr, \
W, Lethbridge, AB T1K 3M4, Canada}
\author{Nicholas swidinsky}
\affiliation{University of Lethbridge,
\ 4401 University Dr, \
W, Lethbridge, AB T1K 3M4, Canada}
\author[0009-0005-2265-2506]{Ryan Wills}
\affiliation{Rochester Institute of Technology, \
1 Lomb Memorial Drive, \
Rochester, NY 14623, USA}
\author[0000-0001-8253-1451]{Michael Zemcov}
\affiliation{Rochester Institute of Technology, \
1 Lomb Memorial Drive, \
Rochester, NY 14623, USA}
\affiliation{Jet Propulsion Laboratory, \
4800 Oak Grove Drive, \
Pasadena, CA 91109, USA}

\author{Derek Arthur}
\affiliation{University of Lethbridge,
\ 4401 University Dr, \
W, Lethbridge, AB T1K 3M4, Canada}
\author[0000-0002-0941-0407]{Victoria Butler}
\affiliation{Cornell University, \
109 Clark Hall, \
Ithaca, NY 14850, USA}
\author[0000-0002-9330-8738]{Richard M. Feder}
\affiliation{Berkeley Center for Cosmological Physics, University of California, Berkeley, CA 94720,
USA}
\affiliation{Lawrence Berkeley National Laboratory, Berkeley, California 94720, USA}
\author[0000-0002-5476-1898]{Daniel Klyde}
\affiliation{Rochester Institute of Technology, \
1 Lomb Memorial Drive, \
Rochester, NY 14623, USA}
\author[0000-0002-3754-2415]{Lorenzo Lovisari}
\affiliation{INAF, Istituto di Astrofisica Spaziale e Fisica Cosmica, Via Alfonso Corti 12, 20133 Milano, Italy}
\affiliation{Center for Astrophysics \text{\textbar} Harvard \& Smithsonian, 60 Garden St., Cambridge, MA 02138, USA}
\author[0000-0002-8031-1217]{Adam Mantz}
\affiliation{Kavli Institute for Particle Astrophysics and Cosmology, Stanford University, 452 Lomita Mall, Stanford, CA 94305, USA}
\author[0000-0002-1616-5649]{Emily M. Silich}
\affiliation{California Institute of Technology, \
1200 E California Blvd, \
Pasadena, CA 91125, USA}

\begin{abstract} 
The conversion of gravitational potential to kinetic energy results in an intracluster medium (ICM) gas with a characteristic temperature near 10 keV in the most massive galaxy clusters. X-ray observations, primarily from \chandra\ and \xmm, have revealed a wealth of information about the thermodynamics of this gas. However, two regimes remain difficult to study with current instruments: superheated gas well above 10~keV generated by shocks from major mergers, and distant systems strongly impacted by cosmological dimming. Relativistic corrections to the Sunyaev-Zel'dovich effect (\RSZe) produce a fractional spectral distortion in the cosmic microwave background at submillimeter and millimeter wavelengths that could offer a complementary probe of both high-temperature and high-redshift ICM gas. Here we describe multiband measurements of the \RSZe, including observations from the Fourier Transform Spectrometer on the \spire\ instrument, that constrain the ICM thermodynamics of the major merger \macs.  Within the seven observed lines of sight, we find an average temperature of \TSZm~$= 15.1\substack{+3.8\\-3.3}$~keV, which is consistent with the values obtained from X-ray measurements of the same regions, with \TChan~$=18.0\substack{+1.1\\-1.1}$~keV and \TXMM~$=13.9\substack{+0.9\\-0.9}$~keV. This work demonstrates that the \RSZe\ signal can be detected with moderate spectral resolution submillimeter data, while also establishing the utility of such measurements for probing superheated regions of the ICM.
\end{abstract}

\section{Introduction} \label{sec:intro}
The intracluster medium (ICM) comprises approximately $15$\% of the total mass and approximately $85$\% of the baryonic mass of galaxy clusters \citep{Gonzalez2007, cluster_mass}, making the ICM an important and dynamic constituent in these systems. Direct observations of the structure and thermodynamics of the ICM are critical to understanding the evolution of clusters and their relationship to the underlying dark matter distribution over cosmic time. In the most massive objects ($M \simeq 10^{15}$~\msun), the ICM can be heated to temperatures near $10^8$~K ($k T \simeq 10$ keV) through conversion from potential to kinetic energy as matter accretes onto the cluster \citep{Kravtsov2012}. Additionally, major mergers drive shocks that can heat regions of the ICM to much higher temperatures, with $kT_{\rm ICM} \simeq 40$ keV inferred in extreme systems \citep[e.g.,][]{Markevitch2006, Markevitch2007}.

The ICM is primarily observed using X-rays where the hot gas emits directly via thermal Bremsstrahlung continuum and line emission \citep[e.g.,][]{Birzan2004,Vikhlinin2006,Markevitch2007,Cavagnolo2009,Arnaud2010}. A spectral fit of this emission permits inference of the electron temperature, $T_e$  \citep[e.g.,][]{Sarazin1986, schellen}. However, X-ray instruments are typically optimized for energy bands below the highest temperatures expected in the ICM, with sensitivity dropping rapidly by $E \simeq 8$~keV for both \xmm\ \citep{xmm_obs} and \chandra\  \citep{chandra_obs}. While not traditionally utilized for ICM studies, X-ray facilities such as \textit{NuStar} do have the capability to observe photon energies up to $50$ keV \citep{nustar}, but challenges related to its broad point spread function (PSF) and small collecting area limit its ability to measure ICM temperatures \citep[e.g.,][]{Wik2014}. As a result of these instrumental limitations, regions of particularly hot gas in major mergers are challenging to characterize via X-rays, which is further compounded by cosmological dimming in high-redshift clusters \citep[e.g.,][]{Calzadilla2024}. There is strong motivation to develop alternative ways to measure the temperature of the ICM.

\begin{figure}
    \centering
    \includegraphics[width=1\linewidth]{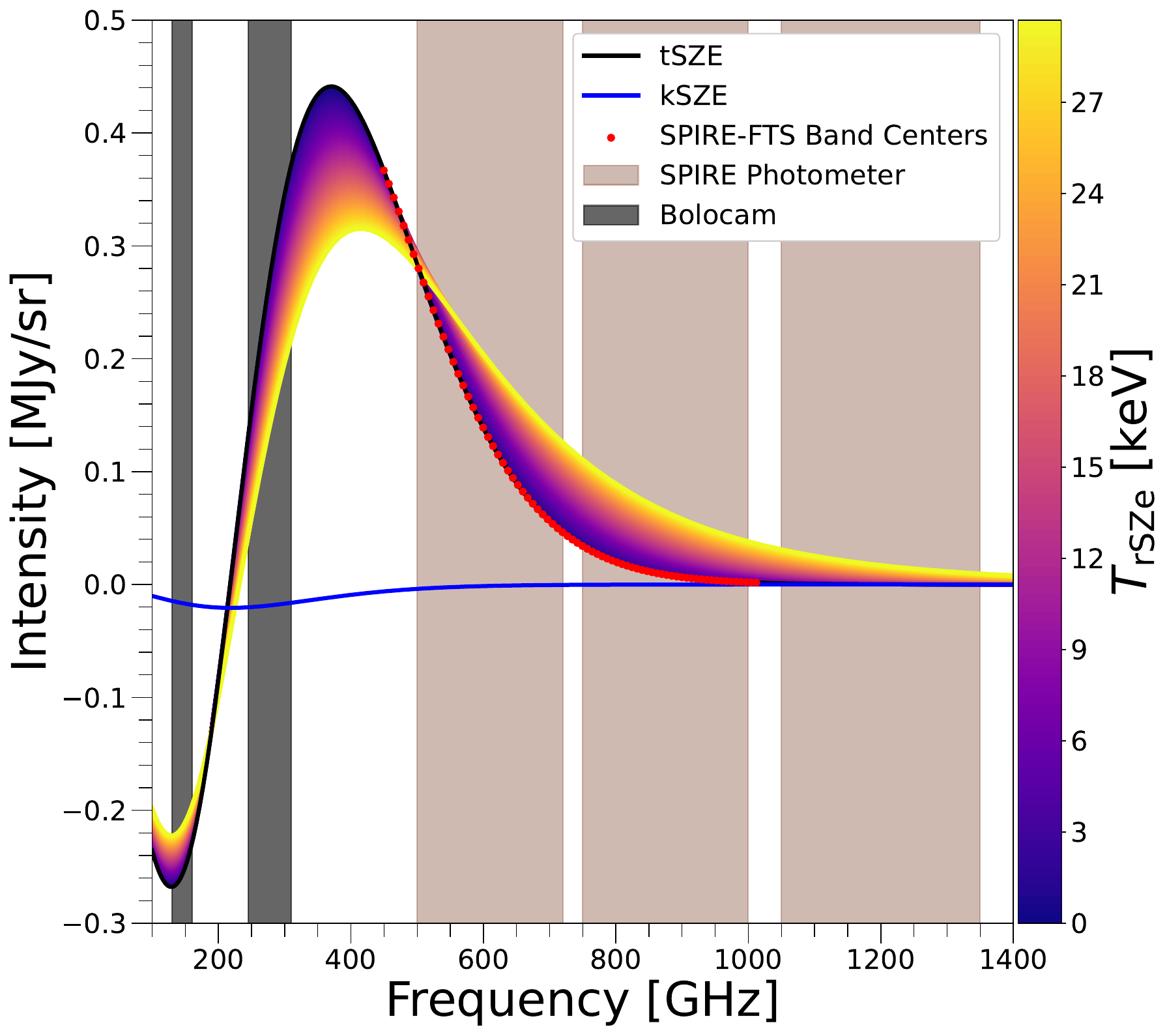}
    \caption{The spectrum of each component of the \SZe\ for a range of \TSZm\ calculated from \szpack\ \citep{SZpack1,SZpack2} assuming a Comptonization $y = 2.4 \times 10^{-4}$ and a peculiar velocity of $800$ km s$^{-1}$. Dark gray regions denote the observing bands of Bolocam \citep{bolo_instrument} and red points indicate the band centers of the \spirefts, both of which are utilized to fit the \SZe\ spectrum in our analysis. Also shown as light gray regions are the observing bands of the \spire\ photometer \citep{observer_manual}, which are used to estimate the contaminating signal from thermal dust emission.}
    \label{fig:sze_theo}
\end{figure}

The Sunyaev-Zeldovich effect (\SZe) offers a probe of ICM thermodynamics that is independent of and complementary to X-ray observations \citep{2019_Sze, rsz_universe}. The SZe is the spectral distortion in the cosmic microwave background (CMB) radiation caused by inverse-Compton scattering of cold $T_{\rm CMB} = 2.73 $ K photons from energetic ICM electrons \citep{sze_1972,Birkinshaw1999}. 
For a stationary plasma, the effect is dependent on the thermal properties of the gas (\TSZe), and can be modeled as an intensity shift in the CMB spectrum described by:
\begin{multline}
    \Delta I_{{\rm \TSZe}} = I_0 \frac{x^4 e^x}{(e^x - 1)^2}\left(x\frac{e^x + 1}{e^x - 1} - 4 \right)\\
    \times(1 + \delta(x,T_e) )\int n_e \frac{k T_e}{m_e c^2}\sigma_T dl,
    \label{eq:TSZe}
\end{multline}
 where $I_0 = \frac{2(k\Tcmb)^3}{(hc)^2}$ and $x \equiv \frac{h\nu}{k\Tcmb}$ \citep{carlstrom}.
The term expressing an integral along the line of sight gives the Compton $y$ parameter, which indicates the average energy boost to a CMB photon as a proportion of the electron rest mass. The ICM is sufficiently hot that  mild relativistic corrections to the classical scattering are required to fully describe the intensity spectrum \citep{Wright1979}. These corrections can be included in the term $\delta(x;T_e)$, a modification that is often referred to as the relativistic correction \citep[\RSZe;][]{Itoh1998, SZpack1, SZpack2, evidence_for_rsz, mm_universe_rsz}. A kinematic effect (\KSZe) is also present due to coherent bulk velocities of and in the ICM material along the line of sight, and is modeled by:
\begin{equation}
    \Delta I_{\KSZe} = -I_0 \beta \tau \frac{x^4e^x}{(e^x-1)^2}.
    \label{eq:KSZe}
\end{equation}
Above, $\beta$ is the ratio of the cluster's peculiar velocity to the speed of light and $\tau$ is the optical depth of electron scattering. The spectra of both the thermal and kinematic effects are plotted in Fig.~\ref{fig:sze_theo}, assuming properties of the ICM typical of massive clusters.  The relativistic corrections for a range of ICM temperatures are also shown, which highlights the temperature dependence of the \RSZe\ corrections. Since the classical (nonrelativistic) \TSZe\ brightness is sensitive solely to the product $n_e T_e$ at a given $x$, it alone cannot be used to isolate the eletron temperature. However, the spectrum of the relativistic correction depends on $T_e$, and thus offers a way to probe electron temperature independent of X-ray observations \citep{mm_universe_rsz}. 

The \RSZe\ is an emerging observable tool that can be used to independently assess the extremely high gas temperatures present in some media, with the shock-heated ICM being an example.  The typical small change in \SZe\ amplitude induced by the correction and its spectral degeneracy with the \TSZe\ and \KSZe\ below 300 GHz has restricted its use in traditional radio- and millimeter-wavelength observations. However, combining sensitive millimeter-wavelength measurements with submillimeter measurements has been demonstrated to be a potentially fruitful way to isolate the \RSZe\ corrections and explore extreme ICM temperatures \citep{Zemcov2010,Zemcov2012,butler}, and recent statistical measurements using \planck\ have also utilized the \RSZe\ as a probe of ICM temperature \citep{Hurier2017,erler,evidence_for_rsz}.

The site of a quadruple-merger, \macs\ is an excellent testbed for measurements of the \RSZe. An analysis of this cluster's morphology \citep{Ma2009} was undertaken via Chandra X-ray measurements and spectroscopy of the cluster-member galaxies, and found the overall merger geometry to be the following. There are two primary systems in the SE region that are undergoing a major merger oriented in the plane of sky, and contain regions of coincident shock-heated ICM gas. In the NW, another subcluster appears to be infalling along the line of sight, but contains significant cool dense gas, thus indicating that it has not yet reached pericenter passage. Further NW is a second subcluster that is largely stripped of its gas, which is suggestive of an advanced merger stage near the apocenter. This is supported by radio observations of the cluster undertaken with the Jansky Very Large Array \citep{Weeren}, where they find $100$-$300$~kpc radio filaments in the outskirts of the cluster that trace the shock-heated gas and that the northwest subcluster is ram pressure stripped. In further support of the \citet{Ma2009} picture, \citet{Sayers2013} and \citet{nika_macs0717} have found large \KSZe\ signals that are coincident with the infalling NW system and with inferred velocities consistent with the cluster-member spectroscopy. Throughout these works, evidence for shock-heated gas is consistently found. In particular, \citet{Breuer} find strong evidence for shock-heated gas near the cluster center, coincident with the \RSZe\ signal, thus making it one of the brightest \RSZe\ targets.

In this work, we describe an analysis wherein observations from the \herschel\ Space Observatory's Spectral and Photometric Imaging REceiver Fourier Transform Spectrometer (\spirefts; \citealt{SPIRE-FTS}) are used to constrain the \RSZe\ signal from the quadruple-merger \macs.  Between $500$ and $900$~GHz the fractional contribution of the \RSZe\ correction is maximized (Fig.~\ref{fig:sze_theo}), and its spectral behavior removes degeneracies with the classical \TSZ\ distortion. These measurements of the \SZe\ spectra with a resolving power $\lambda / \Delta \lambda = R \simeq 30$ from a fully spectrometric instrument provide a space to explore feasibility, assess the effects of various systematic errors, and provide new insights on the state of the cluster's ICM. 

This paper is organized as follows. In Section \ref{sec:data} we describe the primary and ancillary data used in our analysis and how instrument systematics are removed from the data. Section \ref{sec:FTS_reduction} is a detailed presentation of the \spirefts\ data reduction. In Section \ref{sec:fits} we describe the fitting procedure, and in section \ref{sec:discussion} we discuss our results in the context of past and future \RSZe\ measurements.  

\begin{figure*}[ht!]
    \centering
    \includegraphics[width=1\linewidth]{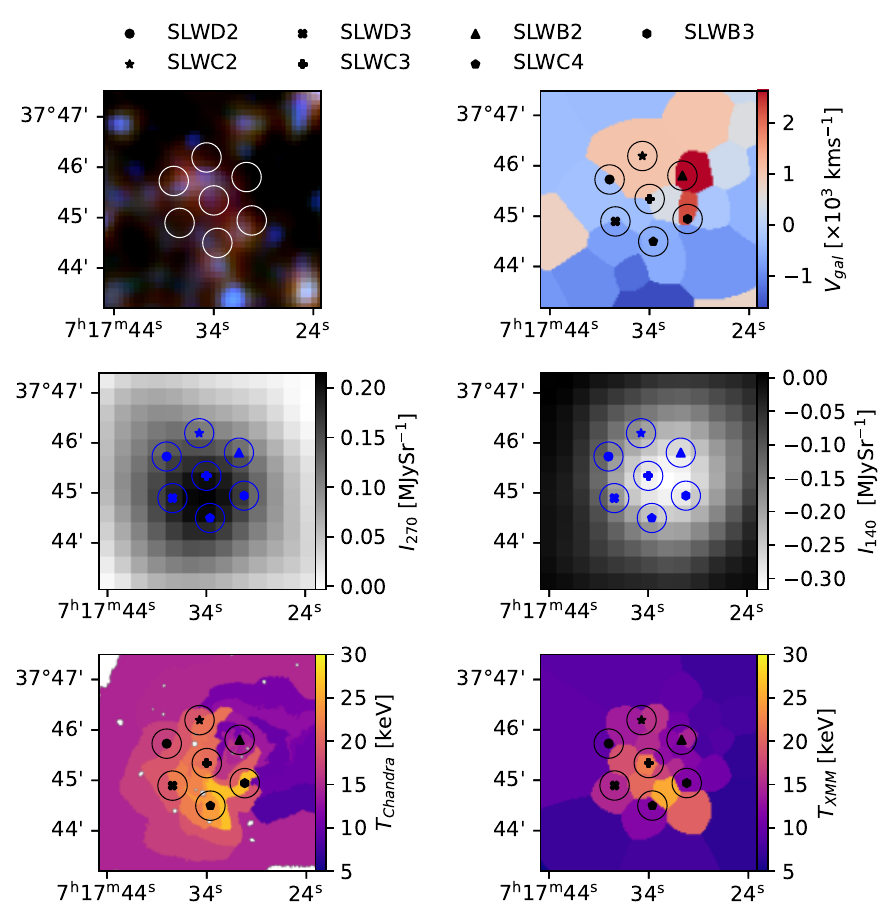}
    \caption{Datasets used in this analysis. The markers (labeled at the top) and circles show the average pointing location and effective beam size for each \spirefts\ SLW detector considered in this work. Top left: false color image of \macs\ from the \spire\ photometer where red represents $600$ GHz emission, green $856$ GHz, and blue $1200$ GHz. Top right: velocity map from cluster-member redshifts.
    Middle row: Bolocam measurements of the \SZe\ intensity at $270$ and $140$ GHz. Bottom row: electron temperature measured by \chandra\ and \xmm.    }
    \label{fig:all-of-the-data}
\end{figure*}

\section{Data} \label{sec:data}

To measure the \RSZe\ in \macs, we combine observations from the \spirefts\ \citep{SPIRE_both} with a variety of ancillary observations at a wide range of wavelengths.  Following reduction of the \spirefts\ measurements to data hypercubes, they are combined with Bolocam \citep{bolo_instrument} intensity measurements to determine the spectrum of the \SZe\ (\S\ref{sec:fits}). Next, \spire\ photometer observations are used to estimate the contaminating signal contributed by dusty infrared galaxies, and cluster-member velocity maps are used as a prior on the ICM velocity \citep{Silich2024}. The corrected intensity spectra are then modeled, and various ICM and instrumental parameters are constrained, including \TSZm.  Finally, the X-ray derived temperatures are combined in a way that makes them comparable to the \RSZe\ measurements. Maps summarizing the data sets are shown in Figure \ref{fig:all-of-the-data}, and a compilation of fiducial values from each of the data products listed above at each position shown in Figure \ref{fig:all-of-the-data} is provided in Table \ref{tab:prior_values}.  In the following section, we present a summary of each source of data.

\subsection{The Herschel SPIRE-FTS}
\label{sec:SPIRE-FTS}

\begin{table*}[ht!]
    \center
    \begin{tabular}{c|c|c|c|c|c|c}
        \hline 
        Observation ID & Spec. Resolution & Spat. Resolution & Target Name & \# of Scans & Date & Integration Time\\
        \hline \hline
        1342252295 & LR & I & \macs & 400 & 2012 Oct 6 & $12,139$ s \\ 
        1342252296 & LR & I & \macs & 400 & 2012 Oct 6 & $12,139$ s\\ 
        1342252297 & LR & SP & Dark Sky & 400 & 2012 Oct 6 & $3,192$ s \\ 
        1342252903 & LR & I& \macs & 400 & 2012 Oct 7 & $12,139$ s \\ 
        1342252904 & LR & I & \macs & 400 & 2012 Oct 7 & $12,139$ s \\ 
        \hline 
    \end{tabular}
    \caption{\spirefts\ observations used in this analysis listed in chronological order by observation date. Intermediate (I) spatial sampling refers to observations with $4$ ``jiggle" positions and single pointing (SP) to one jiggle position within an observation. The number of scans refers to the number of scans taken at a single jiggle position not the total number of scans across the entire observation.}
    \label{tab:obs}
\end{table*}

The \spirefts\ instrument employed a Mach-Zehnder interferometer \citep{MachZ2} as an FTS with two input ports. The first FTS input was illuminated by the \herschel\ telescope, while the second FTS input port was coupled to the Spectrometer Calibrator (SCAL), a cryogenic temperature-controlled source designed to load-match emission from the \herschel\ primary optics \citep{MachZ}. The SPIRE instrument team found that SCAL was not necessary to compensate for the telescope emission during routine observations, but it noticeably increased the photon noise when heated \citep{SCAL_gone}. As a result, SCAL was left to drift with the temperature of the optical bench around $4.5$ K, and for most \spirefts\ observations did not play a role.  Here, we treat the \spirefts\ as a low-resolution (LR) spectro-photometer sensitive to continuum emission; therefore, the SCAL emission will induce a systematic in our data, as discussed further in Section \ref{sec:hipe}.  

The \spirefts\ output focal plane was populated by independent bolometer arrays coupled to the FTS element via a dichroic beamsplitter. The long wavelength array (termed ``SLW'') covered $447-1018$ GHz and comprised $19$ detectors in a hexagonal close-packed pattern, as described in \citet{SPIRE_both}.  There are known difficulties connecting the continuum brightness measured in the short-wavelength array to SLW \citep{Valtchanov2018}, and since the high-frequency data ($944-1568$~GHz) does not constrain the \SZe\ emission in a useful way, we opted to ignore them in this study.

An internal beam steering mirror near the input of the spectrometer permitted spatial modulation of the telescope beam and is used to ``jiggle'' the field of view of the focal plane array to produce fully-sampled images.  The observations under study were performed in the intermediate sampling mode of the instrument, which comprises four jiggle positions separated by one beamwidth \citep{vignett}.  The change in optical configuration was also the source of a strong instrumental systematic, as discussed in Section \ref{sec:hipe}. 

The fundamental \spirefts\ data product is an interferogram acquired over an FTS scan with temporal duration set by the desired spectral resolving power.  The interferograms are processed through the \herschel\ Interactive Processing Environment (HIPE; \citealt{hipe_mandatory, HIPE}). 
Our observations (shown in Table \ref{tab:obs}) are taken in intermediate spatial sampling, $R \sim 30$, LR.  Four hundred interferograms were recorded per jiggle position per observation. 

The standard \spirefts\ observation set included an observation of a ``Dark Sky'' field \citep{dark_field} near the north ecliptic pole used to monitor the characteristics of the instrument using a consistent, faint astrophysical background illumination \citep{characterizing,marchili}. All \spirefts\ observations have a companion Dark Sky observation taken in the same observing mode. However, the companion Dark Sky observation for the \macs\ observations presented here has only one jiggle position, meaning there is no analogue Dark Sky measurement for three of the four optical configurations.  We describe the impacts of this discrepancy in Section \ref{sec:hipe}.

\begin{table*}[t!]
    \centering
    \begin{tabular}{c|c|c|c|c|c|c|c}
         \hline 
         Detector & SLWD2 & SLWC2 & SLWD3 & SLWC3 & SLWB2 & SLWC4 & SLWB3  \\ \hline \hline 
         \vpec\ [km s$^{-1}$] & $\phantom{00}260\substack{+190\\-190}$ & $\phantom{00}930\substack{+300\\-300}$ & $ -410\substack{+110 \\-110}$ & $\phantom{00}370\substack{+150\\-150}$ & $\phantom{0}1500\substack{+180\\-180}$ & $-540\substack{+90\phantom{0}\\-90\phantom{0}}$ & $\phantom{00}250\substack{+110\\-110}$  \\ 

         \TXMM\ [keV] & $\phantom{0}12.3\substack{+0.4\\-0.4}$ & $\phantom{0}12.8\substack{+0.4\\-0.4}$ & $\phantom{0}14.1\substack{+0.5\\-0.5}$ & $\phantom{0}16.9\substack{+0.4\\-0.4}$ & $\phantom{0}11.7\substack{+0.2\\-0.2}$ & $\phantom{0}15.7\substack{+0.5\\-0.5}$ & $\phantom{0}14.3\substack{+0.4\\-0.4}$ 
         \\ 
         \TChan\ [keV] & $\phantom{0}17.6\substack{+0.8\\-0.8}$ & $\phantom{0}17.0\substack{+0.7\\-0.7}$ &$\phantom{0}19.6\substack{+0.8
         \\-0.8}$ & $\phantom{0}19.7\substack{+0.6\\-0.6}$ & $\phantom{0}14.2\substack{+0.3\\-0.3}$ & $\phantom{0}21.7\substack{+0.9\\-0.9}$ & $\phantom{0}17.2\substack{+0.7\\-0.7}$ 
         \\
         $I_{140}$ [MJy sr$^{-1}$]& $-0.16\substack{+0.02\\-0.02}$ & $-0.20\substack{+0.02\\-0.02}$ & $-0.18\substack{+0.02\\-0.02}$ & $-0.31\substack{+0.02\\-0.02}$ & $-0.29\substack{+0.02\\-0.02}$ & $-0.23\substack{+0.02\\-0.02}$ & $-0.30\substack{+0.02\\-0.02}$ 
         \\ 
         $I_{270}$ [MJy sr$^{-1}$]& $\phantom{-}0.15\substack{+0.02\\-0.02}$ & $\phantom{-}0.12\substack{+0.02\\-0.02}$ & $\phantom{-}0.18\substack{+0.02\\-0.02}$ & $\phantom{-}0.21\substack{+0.02\\-0.02}$ & $\phantom{-}0.13\substack{+0.02\\-0.02}$ & $\phantom{-}0.20\substack{+0.02\\-0.02}$ & $\phantom{-}0.16\substack{+0.02\\-0.02}$ 
         \\
        R.A. & 07\RAhour17\RAmin38\RAsec01 & 7\RAhour17\RAmin34\RAsec72 & 7\RAhour17\RAmin37\RAsec43 & 7\RAhour17\RAmin33\RAsec99 & 7\RAhour17\RAmin30\RAsec70 & 7\RAhour17\RAmin33\RAsec63 & 7\RAhour17\RAmin30\RAsec18 
        \\ \
        Decl. & 37\degree45\arcmin43\declsec7 & 37\degree46\arcmin11\declsec68 & 37\degree44\arcmin53\declsec77 & 37\degree45\arcmin20\declsec52 & 37\degree45\arcmin48\declsec36 & 37\degree44\arcmin29\declsec91 & 37\degree44\arcmin56\declsec7 
        \\ \hline
         
    \end{tabular}
    \caption{Parameter values obtained from the auxiliary data for each of the seven \spirefts\ pointing positions shown in Figure \ref{fig:all-of-the-data}. The prior on ICM velocity utilized in our fits, \vpec\ is derived from data in Section \ref{sec:velocity_data} and described in Section \ref{sec:fits}. Electron temperatures from \xmm\ and \chandra\ are derived from the data described in Sections \ref{sec:chandra} and \ref{sec:XMM} using the algorithm presented in Section \S\ref{sec:data_extract}. The 140 and 270~GHz intensities were obtained from Bolocam as detailed in Section \ref{sec:bolocam}. }
    \label{tab:prior_values}
\end{table*}

\subsection{SPIRE Photometer}

The \spire\ instrument also contains a set of three photometric channels, which are henceforth referred to as the \spire\ photometer (SPIRE-P, \citet{Griffin_2006}). These channels have band centers of $600$, $856$, and $1200$~GHz, with beam sizes of $36\arcsec$, $25\arcsec$, and $18\arcsec$, respectively. The photodetectors in use are the same kind of bolometers utilized in the SPIRE-FTS and are arranged in three grids of $43$, $88$, and $139$, which are illuminated by the same optics as the \spirefts, though the two instruments are mutually exclusive, with only one able to observe at a time. Each of these arrays have a subset of $10$ detectors that are coaligned allowing for easy comparison between the three channels. The photometer has a field of view of $4\arcmin \times 8\arcmin$. During observations, a two-dimensional raster scanning strategy is put into service, where the focal plane is rotated $\pm42.4\degree$ to account for the underfilled detector arrays \citep{observer_manual}. We use a SPIRE-P observation to constrain the emission of the cosmic infrared background (CIB) and reduce it from our data, which is described in more detail in \S\ref{sec:CIB}.

\subsection{Bolocam}
\label{sec:bolocam}

Bolocam was used to observe \macs\ in two different photometric bands centered on approximately $140$ and $270$ GHz. The data are reduced to surface brightness maps using the methods described in \citet{Sayers2013} and \citet{Sayers2019}. Here, we make use of these surface brightness images in both bands, which both readily cover the field of view of the \spirefts\ with approximately uniform depth. The Bolocam map uncertainty is assumed to follow a Gaussian distribution with a standard deviation estimated from extracting the surface brightness from identical locations in $1000$ random noise realizations of the data. The FWHM of the Bolocam PSF, equal to 59\arcsec\ at 140~GHz and 33\arcsec\ at 270~GHz, is comparable to that of the \spirefts\ over the frequency range considered in our analysis. Therefore, we do not attempt to correct for any PSF effects in our surface brightness measurements. The precise values of $I_{270}$ and $I_{140}$ used throughout this work are shown in Table \ref{tab:prior_values}.

\subsection{cluster-member Redshifts}
\label{sec:velocity_data}
In the Bolocam bands, the \KSZe\ due to the line-of-sight velocity of the gas can produce a nonnegligible signal compared to the \RSZe\ signal we aim to measure, as shown in Figure \ref{fig:sze_theo}. Due to strong degeneracies between the amplitudes of the \KSZe\ and \TSZe\ signals, along with the gas temperature (which specifies the relativistic corrections to those signals), it is not possible to obtain meaningful constraints on all three parameters with the existing spectral coverage. As demonstrated in \citet{Sayers2013} the line-of-sight gas velocities are consistent with the bulk dark matter velocities estimated from spectroscopy of cluster-member galaxies \citep{Ma2009}. We therefore use external data to place a prior on the gas velocity in the region of the cluster observed by the \spirefts.

Following the technique of \citet{Silich2024}, we obtain a map of the dark matter velocity by spatially binning the cluster-member galaxy spectra. To summarize the analysis, we first define a spatial aperture based on the cluster-member galaxy distribution by convolving a cluster-member galaxy count map constructed via the redshift catalog of \citet{Ma2009} with a $2$\arcmin\ peak-normalized top-hat kernel. This new map is then smoothed via convolution with a $3$\arcmin\ FWHM area-normalized Gaussian kernel. We then draw a contour within which each pixel has $\gtrsim 5$ galaxies. To close the contour at large radii, we define a circular aperture with $r = 6.5'$ ($\simeq2.5$ Mpc at $z \simeq 0.5$). At any point where the drawn contour would remain open at this circular boundary, we close the contour with the circular aperture. Within this aperture, we generate the spatial bins for the cluster-member galaxy velocity map by applying a publicly-available Python implementation of the weighted Voronoi tessellation (WVT) algorithm \citep{Diehl2006,rhea2020} to our galaxy counts map, requiring $10$ galaxies per spatial bin to ensure statistical significance and consistency between pixels. We then apply an estimator from \cite{Beers1990} to populate the cluster-member galaxy velocity in each bin.

\subsection{X-ray Spectroscopy}
\label{sec:x-ray data}
We derive X-ray inferred gas temperatures from both \chandra\ and \xmm\ to compare with our derived values of \TSZm\ following the data processing techniques described below.
\subsubsection{Chandra}
\label{sec:chandra}

The reduction and analysis of the Chandra data are described in detail by \citet{new_mantz}, using {\sc ciao} version 4.16 and {\sc caldb} version 4.11.0.
In brief, the archival data (Observation 1655, \citet{chandra_data}) were reprocessed using standard techniques to produce level 2 event files and filtered to remove periods of enhanced background.
The source detection code {\sc wavdetect} was used to find pointlike sources, which were masked in later analysis.
Images were produced in the 0.6--7.0\,keV band, binned to 0.984$''$ resolution, and regions for spectral analysis were defined using the {\sc contbin} algorithm \citep{contbin} with a target signal-to-noise ratio (S/N) of 55 and a geometric constraint parameter of 2.
Spectra in these regions were modeled as the sum of a Galactic soft foreground, constrained by ROSAT All-Sky Survey data \citep{sabol2019}; emission from active galactic nuclei fainter than the  {\sc wavdetect} detection limit; the quiescent particle-induced background \citep{suzuki2021}; and absorbed thermal emission from the ICM, using the {\sc apec} code and adopting an equivalent absorbing hydrogen column density of $N_H=6.92\times10^{20}$\,cm$^{-2}$ \citep{h1p_collab}.
Full details of the foreground and background modeling, including systematic allowances marginalized over, can be found in \citet{new_mantz}.
All spectra were fit in the 0.6--7.0\,keV observer-frame band using {\sc xspec} (version 12.12.1c) to evaluate the likelihood and the {\sc lmc} python package to sample the posterior distribution.
The analysis assumed a common ICM metallicity among all regions, allowing the temperatures and normalizations to fit independently.
The resulting map of best-fitting temperatures is shown in Figure \ref{fig:all-of-the-data}.

\subsubsection{XMM-Newton}
\label{sec:XMM}
We use the European Photon Imaging Camera to observe this cluster with \xmm. This instrument uses two types of cameras: pn CCDs and MOS CCDs. We use observation $0672420201$, and the data files were reprocessed using the \xmm\ Science Analysis System v19.1.0. 
We employed the tasks {\it emchain} and {\it epchain} to generate calibrated event files from the raw data. 
For the pn camera, an out-of-time event file was also produced using {\it epchain}.
To ensure data quality, we excluded all events with PATTERN$>$12 for the MOS cameras and PATTERN$>$4 for the pn camera. Additionally, events adjacent to CCD edges and those associated with bad pixels were removed by applying FLAG = 0 for all cameras. 
We also excluded data from CCDs identified as being in an anomalous state, following the criteria described by \cite{2008A&A...478..575K}.
Finally, periods of high background were filtered out using the {\it mos-filter} and {\it pn-filter} tasks.

Spectral analysis was carried out using the XSPEC package \citep{1996ASPC..101...17A}, version 12.12.0. 
The energy ranges considered were 0.5–12 keV for the MOS cameras and 0.5–14 keV for the pn camera. 
The X-ray emission was modeled with an APEC thermal plasma model \citep{2001ApJ...556L..91S}, adopting elemental abundances from \cite{2009ARA&A..47..481A} and employing C-statistics for the spectral fitting.

Photoelectric absorption was fixed to the total hydrogen column density, including both neutral and molecular components, using the value $N_H$=8.36$\times10^{20}$ cm$^{-2}$, as reported by \cite{2013MNRAS.431..394W}. 
The MOS and pn spectra were fit simultaneously, with the temperature and metallicity parameters linked across instruments while allowing the normalizations to vary independently.
The \xmm\ background modeling is rather complicated, and we defer the interested reader to \cite{2019MNRAS.483..540L} for a detailed description.

The temperature map shown in Figure \ref{fig:all-of-the-data}  was produced following the methodology described in \cite{chex}. 
Spatial regions were defined using the WVT binning algorithm \citep{Diehl2006}, with an S/N requirement of $\ge$ 50.

\subsection{Extracting Fiducial Values from Tessellated Maps}
\label{sec:data_extract}

The maps of cluster-member galaxy velocities and the electron temperatures from both \chandra\ and \xmm\ are composed of irregular-shaped regions defined by the WVT or \texttt{contbin} procedures. To extract the fiducial values of the velocity fields at each of the \spirefts\ pointings we use a PSF-weighted average computed from the entire map. The weighting factor is given by $\frac{B}{\sigma^2}$ where $B$ is the fraction of the PSF that couples to a given region and $\sigma$ is the measurement uncertainty on the quantity within that region. The \spirefts\ SLW bands' PSF depends on frequency channel, but we make the approximation that it has a Gaussian shape with a FWHM equal to 35\arcsec\ for all channels. For comparison, the range of PSF sizes over the frequencies used in this analysis are $31\arcsec$ to $37\arcsec$ with an average of $34\arcsec$. Conversely, for the temperature maps, we opt to use a per-pixel bin weighting because of the fact that the X-ray \texttt{contbins} are chosen to have an approximately equal number of total counts, and thus an inverse variance weighted approach would be biased toward lower temperatures. The extracted temperature values and their uncertainties are estimated via a Monte Carlo sampler. We produce $10^4$ realizations of the temperature maps, where each spatial bin is varied as a Gaussian-distributed draw with mean and standard deviation set by the central value and measurement uncertainty. The fiducial temperatures are then taken to be the median and standard deviation of this set of $10^4$ realizations and are shown, alongside the fiducial velocity values, in Table \ref{tab:prior_values}.

\section{SPIRE-FTS Data Reduction}
\label{sec:FTS_reduction}

To produce submillimeter spectra that can be compared against models for the \RSZe, the raw interferograms reported by the \spirefts\ are corrected for detector and optical effects, calibrated, and mapped onto spectral cubes.  Systematic instrumental effects are then modeled and removed.  The resulting spectral hypercubes are then corrected for foreground astrophysical emission to produce estimates of the diffuse \SZe\ signal.  

\subsection{Spectral Intensity Estimation and Data Cuts}
\label{sec:hipe}
\begin{figure}
    \centering
    \includegraphics[width=1\linewidth]{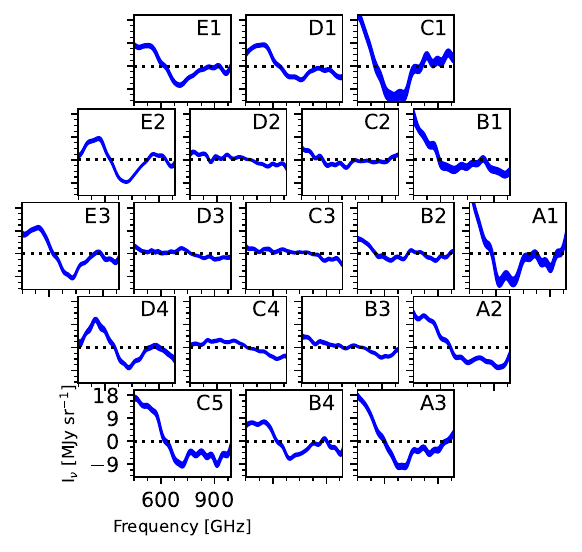}
    \caption{The Dark Sky observation described in Table \ref{tab:obs} as processed through HIPE following the procedure outlined in \S\ref{sec:hipe} including the standard LR correction. The widths of the lines represent the per-datum error estimate from HIPE. All panels have the same $x$ and $y$ axes, which are labeled only for the bottom left panel. A systematic pattern associated with beam vignetting is evident in the outer ring of detectors; for example, the average standard deviation of the intensity of these outer ring detectors is $6.3$~MJy sr$^{-1}$ compared to $1.9$~MJy sr$^{-1}$ for the inner detectors. There remains a low-frequency increase up to $\sim 6$~MJy sr$^{-1}$ that is consistent with a ~$4.5$~K blackbody, as we would expect from the beam intercepting cold optics with slowly varying temperatures.}
    \label{fig:vignette_lr}
\end{figure}

The HIPE Spectrometer Mapping User Reprocessing Script v.14 is used to generate a spectrum of \macs\ at each of the 112 pointing positions (7 detectors $\times$ 4 observations $\times$ 4 jiggle positions) using the v14.3 \spirefts\ calibration tree. A threshold of 4 is used for the baseline correction, and Level 2 deglitching is performed with the {\sc mad} algorithm. 
The \textit{spectrum2d} object is made with beam diameters estimated from SLWC3 and the extended source calibration correction is applied.  In Figure \ref{fig:vignette_lr} we present calibrated spectra produced using the HIPE pipeline with the parameters described above, including the standard LR-mode correction. Notably, the deviation of the outer ring detectors is about $3{-}4 \times$ that for the central detectors, and exhibits strongly correlated low-frequency variation over the passband. This is due to a known vignetting effect in the instrument \citep{vignett} that we find is too severe to model adequately.  We therefore opt to perform our analysis on only the seven central detectors where the vignetting is not present, and the \SZe\ signal is largest.
 
\begin{figure}[h!]
    \centering
    \includegraphics[width=1\linewidth]{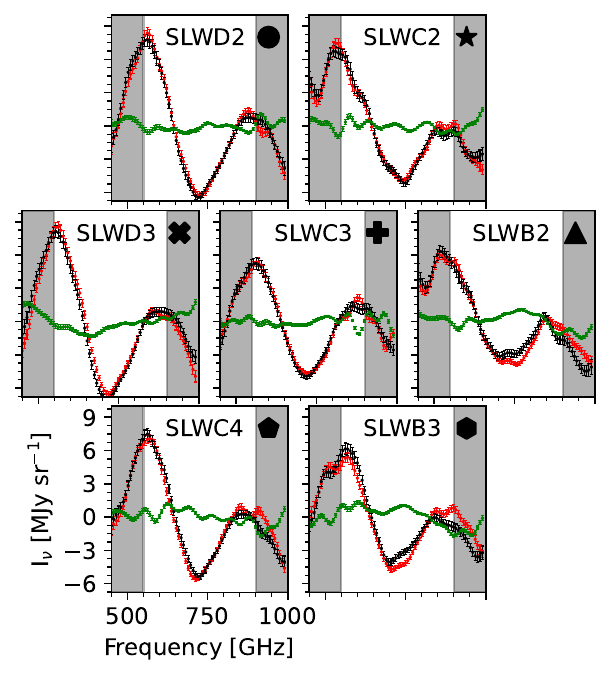}
    \caption{\macs\ spectra after HIPE processing for each of the seven inner SLW bolometers. The black points denote the \spirefts\ spectrum that we retrieve from HIPE, the red points indicate the same for the Dark Sky observations, and the green points indicate the cleaned spectrum obtained from subtracting a scaled version of the Dark Sky data. All panels share the same $x$ and $y$ axes which is depicted only for the lower-left subplot. We estimate the noise covariance of these cleaned data from Equation~\ref{eq:cov}, and the square root of the diagonal elements are shown as error bars. 
    }
    \label{fig:dark_template_subtraction}
\end{figure}

An LR-mode correction implemented as an empirical model \citep{marchili} is normally used to correct an aliased instrumental signal in HIPE. The signal itself appears as a ``double bump'' in the continuum that we refer to as the Marchili signal. The spectra in Fig.~\ref{fig:vignette_lr} exhibit residual Marchili-like structure after processing through the standard HIPE pipeline, which we identify as arising from the generalized empirical approach developed in \citet{marchili} not sufficiently subtracting the effect seen in the observed data. To improve the modeling of systematics, we turn off the standard LR-mode correction step in HIPE and instead use the Dark Sky observation to create a template for subtraction. This is motivated by the fact that the astrophysical signal in the Dark Sky observations is small, and that they are taken in the same observing mode and during identical thermal periods as each of our four observations. They are thus a reliable tracer of the instrumental systematics in a time-adjacent sense, and provide a better template for subtracting the aliased Marchili signal.

The process for creating the Dark Sky templates from each detector spectrum proceeds as follows.
For each detector $i$ and pointing position $j$, an amplitude scaling is computed according to $A_{ij} = \frac{\vec{s}_{ij} \cdot \vec{d}_i}{\vec{d_i} \cdot \vec{d}_i} \vec{d}_i$, where $\vec{d}_i$ is a Dark Sky spectrum and $\vec{s}_i$ is a spectrum of \macs. The corrected spectrum is then computed from $\vec{s}_{{\rm c},ij} = \vec{s}_{ij} - A_{ij} \vec{d}_i$.
The $\vec{s}$, $A \vec{d}$ and $\vec{s}_{\rm c}$ for each detector are shown in Figure \ref{fig:dark_template_subtraction}. For the jiggle position 0 data, this template subtraction leaves a residual with a standard deviation of $0.6$~MJySr${^{-1}}$, approximately a factor of $3$ better than what was achieved using the standard correction and with no discernible Marchili-like structure.  However, this correction works less well for the three jiggle positions that were not present in the Dark Sky observation, which constitute different optical paths through the instrument, and leaves residual Marchili-like structure due to incomplete cancellation.  We therefore opt to exclude data taken in jiggle positions $\{1, 2, 3 \}$ from analysis, which increases the statistical uncertainty by a factor of $2$ but removes a likely uncharacteristic systematic error from the measurement.

\begin{figure}
    \centering
    \includegraphics[width=1\linewidth]{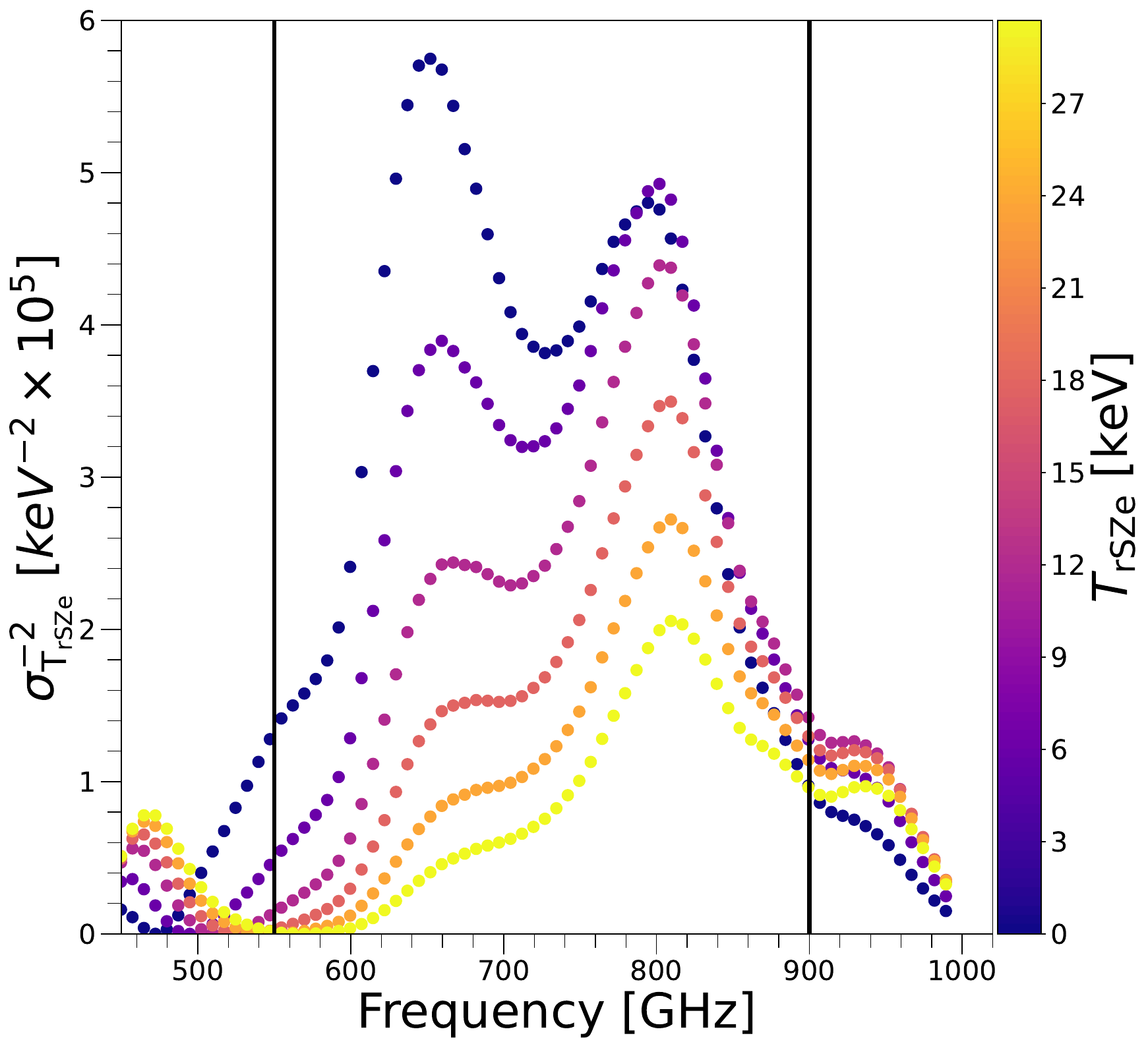}
    \caption{constraining power on \TSZm\  estimated by the inverse square of equation \ref{eq:sze_sensivity}. The solid black vertical lines highlight where the low- and high-frequency cuts from \S\ref{sec:hipe} are applied, and outside of which the constraining power is small. 
    While the \SZe\ distortion is spectrally smooth (see Fig.~\ref{fig:sze_theo}), the relativistic corrections exhibit $\sim 100$ GHz wide spectral deviations whose relative amplitudes change over the falling edge of the SZ spectrum.  These differential effects provide a lever arm to distinguish gas at different temperatures independently of X-ray information. Hence, the constraining power peaks at ~$640$ and ~$800$~GHz. This is informative for cutting the low- and high-frequency ends of the FTS bandpass, as the regions with very low constraining power, e.g. $\nu < 550$, will contribute disproportionately more noise to the measurement. }
    \label{fig:sze_sensitivity}
\end{figure}

Figure \ref{fig:dark_template_subtraction} illustrates a remaining mismatch between the data and Dark Sky template at frequencies near the band edges, again likely arising from instrumental systematics. To produce a calculable figure of merit to inform which spectral information to exclude, we estimate the per spectral channel constraining power as follows. We model the \RSZe\ spectrum using fiducial values from Bolocam measurements and with varying temperature. The constraining power on \TSZm\ is then calculated from:
\begin{equation}
    \sigma_{\rm T}^{2} = \left( \frac{\sigma_I}{\delta I_{\rm T}} \right)^{2}.
    \label{eq:sze_sensivity}
\end{equation}
Here, $\sigma_I$ is the per spectral channel uncertainty, and $\delta I_{\rm T}$ is the partial derivative with respect to \TSZm. The inverse-squared of this is plotted as a function of spectral channel, and for several different values of \TSZm, in Figure \ref{fig:sze_sensitivity}. We conclude the optimal spectral region is 550--900 GHz, as the sensitivity to the \RSZe\ signal outside of this range is small, and including those data would significantly increase the measurement error.  This range also removes the portions of the spectra that may contain significant instrumental systematics.

\subsection{Spectral Noise Estimation}
\label{sec:noise_estimate}
 
\begin{figure}[h!]
    \centering
    \includegraphics[width=1\linewidth]{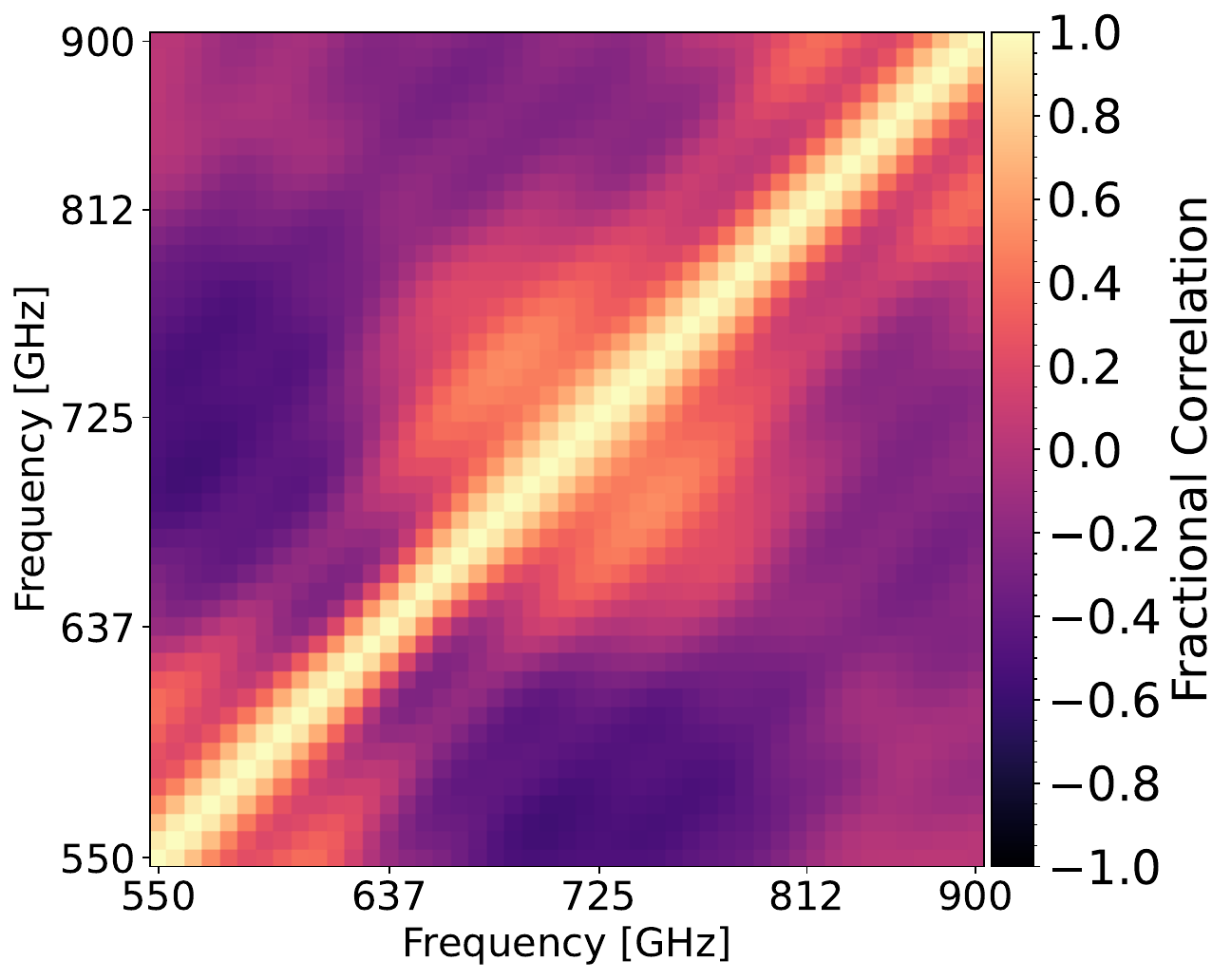}
    \caption{A representative noise correlation matrix for a single \spirefts\ detector. This is computed from the noise covariance matrix $C$ defined in eq \ref{eq:cov}, as $D^{-1}CD^{-1}$, where $D$ is a matrix containing only the square root of the diagonal elements of $C$. The effective correlation length is of order 50~GHz.
    }
    \label{fig:correlation_matrix}
\end{figure}

A covariance matrix is estimated using a timewise jackknife resampling that is designed to cancel spatially and spectrally stable signals in our data, e.g., from astrophysical sources or instrument thermal emission, while retaining variations due to statistical noise and properties of the instrument varying on the scan time scale. To generate this estimate, two hundred of the four hundred spectra for a given pointing are multiplied by negative 1, and the mean is computed with:
\begin{equation}
\vec{s}_{l}^* = \frac{\sum_k \vec{s}_{k\pm}}{2N_{scan}}.    
\end{equation}
Here, $\vec{s}^*_{k\pm}$ denotes the sum over all $400$ spectra for a given pointing, where indices with a minus sign are multiplied by negative one and indices with a plus sign are kept the same. The resulting covariance matrix is estimated by computing the outer product of $\vec{s}_l$ with itself,
\begin{equation}
    C_{l} = \vec{s}_{l}^*\vec{s}_{l}^{\rm T}
    \label{eq:cov}
\end{equation} 
where ${\rm T}$ denotes the matrix transpose. This process is repeated $2 \times 10^6$ times (chosen to estimate the covariance matrices to five significant figures) and averaged. We perform these jackknives independently for each detector, jiggle position (only jiggle position $0$ is used in the analysis), and observation, producing $28$ unique covariance matrices. Since we expect multiple observations with the same detector to have consistent systematics, we average these by detector down to a total of seven. A representative example showing the correlation matrix, which is estimated from the noise covariance matrix, is shown in Figure \ref{fig:correlation_matrix}. The characteristic correlation length implied by the noise covariance is approximately 50~GHz, comparable to the nominal LR spectral resolving power of 25 GHz.

\subsection{Unwanted Astrophysical Emission}
\label{sec:CIB}

The largest source of astrophysical emission toward \macs\ is from the dusty infrared galaxies that make up the CIB. The thermal dust emission from these galaxies peaks near 100 $\mu$m and is redshifted to fall in the \spire\ bandpass \citep{planck_2011}. The CIB signal is much bluer than the \SZe\ signal \citep{erler,planck_dust,butler}, and is brightest where the \SZe\ intensity is negligible above 1 THz. As a result, it is possible to estimate the contribution of the CIB to the \spirefts\ observations using \spire\ photometer data. We compute a model for the CIB galaxy emission in the cluster in three steps. First, we use the Probabilistic Cataloging in the Presence of Diffuse Emission \citep[PCAT-DE;][]{PCAT_2023} software to measure the flux densities and positions of pointlike sources coincident with the galaxy cluster using data from the three \spire\ photometer bands (\S\ref{sec:pcat-fits}). Next, we fit a graybody spectrum to the spectral energy distribution (SED) of each recovered source (\S\ref{sec:graybody-model}). Lastly, we create mock CIB maps at each \spirefts\ frequency to infer the contribution of the CIB to the FTS Observations (\S\ref{sec:graybody-model}) and subtract this model from the spectral data.  Each of these steps is described below.

\subsubsection{Source Catalog Generation} 
\label{sec:pcat-fits}

PCAT-DE uses a Bayesian hierarchical model to jointly fit the flux density and position of pointlike sources, the fluctuation and amplitude of diffuse foreground/background components, and the amplitude of surface brightness templates over a set of multiband photometric maps \citep{PCAT_2023}. In this analysis, we use PCAT-DE to recover the point source SEDs from the \spire\ photometer maps of \macs. 

To construct a point source catalog, \macs\ photometer images in all three \spire\ bands (600, 856, and 1200 GHz) are modeled using a modified version of the procedure developed for \citet{butler}.  Because our goal is to recover a model of the sources comprising the CIB, we fix the \SZe\ template amplitudes for the photometer medium wavelength (PMW)  and photometer long wavelength (PLW) bands to their fiducial values from Bolocam measurements of the optical depth \citep{Sayers2019} and the X-ray-inferred temperature from \xmm\ (\S\ref{sec:x-ray data}) so that the \SZe\ is no longer a free parameter.  Point sources, the cirrus foreground, and an overall offset are then fit over a $10^\prime\times10^\prime$ region centered on the X-ray centroid of the cluster, as described in \citet{PCAT_2023}. This results in a PCAT source catalog describing the positions and fluxes of point sources that is decoupled from diffuse sources of emission in the cluster field.

\begin{figure}[th!]
    \includegraphics[width=1\linewidth]{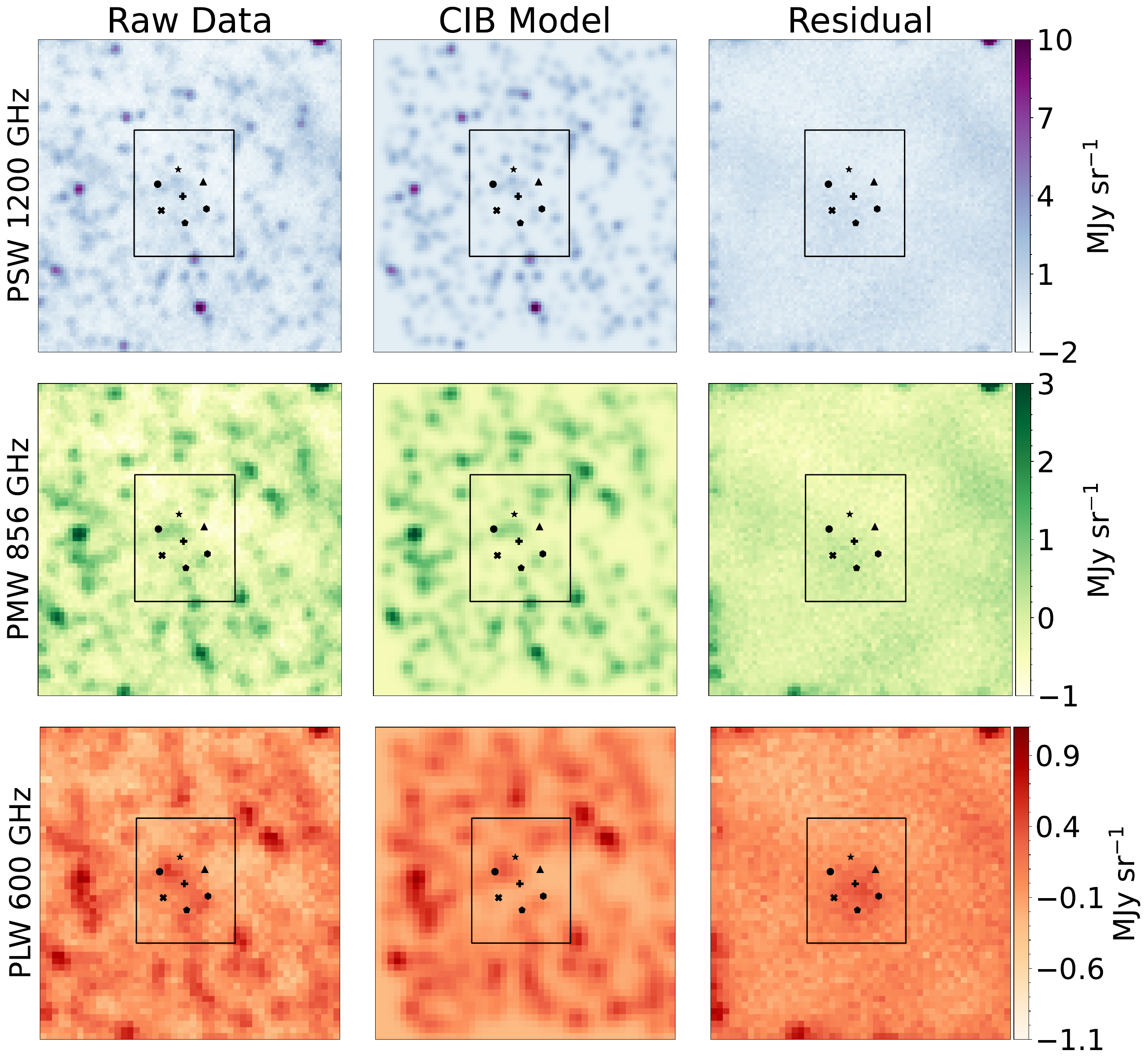}
    \includegraphics[width=0.95\linewidth]{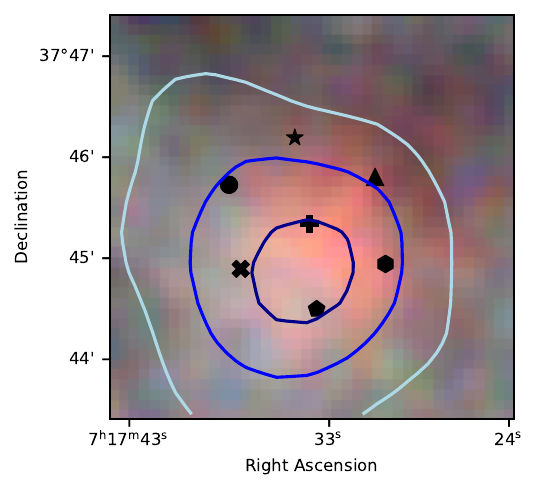}
    \caption{\spire\ photometric imaging of \macs. From top to bottom, the rows show PSW ($1200$ GHz), PMW ($856$ GHz), and PLW ($600$ GHz). From left to right, the columns show the observed image centered on \macs, a fitted CIB point source model constructed using PCAT-DE \citep{PCATII}, and the residual after subtraction of this model. The pointing of each of the seven detectors is shown with their respective symbols as indicated in Figures \ref{fig:all-of-the-data},\ref{fig:dark_template_subtraction},\ref{fig:CIB}, and \ref{fig:sze_spectral_fits}. This demonstrates that we are able to accurately remove CIB contamination while preserving the \SZe\ signal, as primarily seen in the $600$~GHz panel. The bottom panel is a false color image of the three residuals with contours overlaid from the Bolocam $140$~GHz measurements. This illustrates our high-S/N measurement of a diffuse signal that is consistent with the expected red spectrum of the \SZe.
.
    }
    \label{fig:pcat_resids}
\end{figure}

Residuals between the input \spire\ photometer maps and the CIB point source model constructed by \textit{PCAT} are shown in Figure \ref{fig:pcat_resids}.  The CIB model removes point source emission from all three bands and leaves a strong diffuse signal that is brightest in the $600$~GHz band, consistent with the \SZe. A false three-color image of the residuals that can be directly compared to the upper left panel of Figure \ref{fig:all-of-the-data} is also shown in Fig.~\ref{fig:pcat_resids}.  This visually illustrates our high-S/N measurement of diffuse \SZe\ emission in \macs\ with the \spire\ photometer.  

\subsubsection{CIB Correction}
\label{sec:graybody-model}

With a reliable point source catalog, we can next model the spectrum of each point source. A graybody spectrum \citep[e.g.,][]{planck_2013} is fit to the SED of each photometer source identified by PCAT,
\begin{equation}
    I_{\nu} = \frac{\epsilon}{B(\nu_{p}, \Tdust)}B(\nu,\Tdust)\left(\frac{\nu}{\nu_{p}}\right)^{\beta},
\end{equation}
where the free parameter $\epsilon$ describes the amplitude of the source in mJy and $\Tdust$ the temperature of the dust. We set the pivot frequency at $\nu_{p} = 856$ GHz and fix $\beta = 1.65$ to match the \textit{Planck} model \citep{planck_dust}. The fits are implemented via a Monte Carlo Markov Chain (MCMC) sampler \citep{emcee} with loss function given by the $\chi^2$ distribution. Uniform priors of $\Tdust \in [5,200]$~K and $\epsilon \in [0,10^5]$ mJy ensure that these fits remain within a physically meaningful space of values.  The fit outputs are posterior distributions for $\epsilon$ and $T$ for each of the PCAT-detected sources in the cluster region. On a source-by-source basis, SEDs from these posterior predictive distributions are sampled in the \spirefts\ spectral binning, which allows us to create a probability distribution of flux densities for each source $\{F_{\nu,s}\}$.

\begin{figure}
    \centering
    \includegraphics[width=1\linewidth]{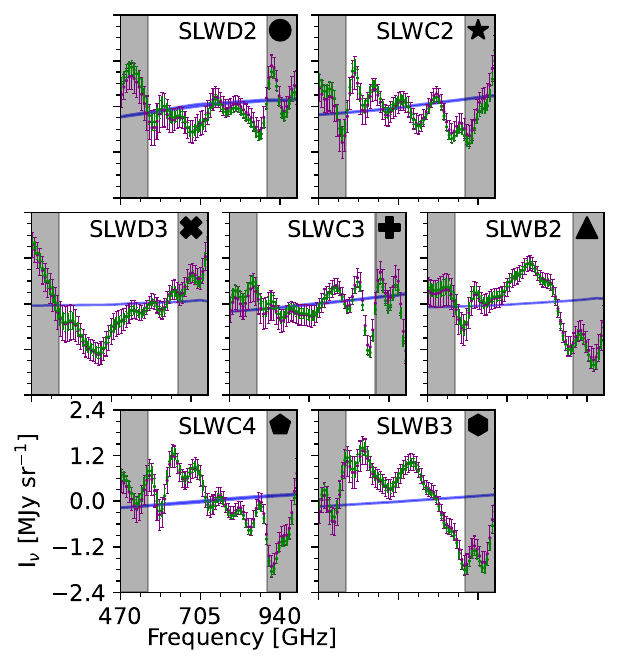}
    \caption{Residuals of the measured spectra and CIB model for each of the seven detectors used in this analysis. The purple points show the Dark Sky template-subtracted signal shown in Fig.~\ref{fig:dark_template_subtraction}, the blue curve is the estimated CIB spectrum with width set by the 68\% model uncertainty, and the green points are the CIB-subtracted measurements. All panels share the same $x$ and $y$ axes, which are depicted only for the lower-left subplot. The CIB corrections act to preferentially reduce flux at high frequencies, as one would expect from the typical graybody model of a dusty infrared galaxy at this redshift. 
    }
    \label{fig:CIB}
\end{figure}

From these probability densities, we generate mock catalogs and project them into corresponding mock images. We create a set of 5,000 such mock catalogs to better sample the source-to-source covariance derived from the PCAT catalog posterior. For each catalog, we assume that each source is pointlike and convolve with the frequency-dependent \spirefts\ PSF to obtain a mock map. Each source's location is fixed to the median position obtained from PCAT-DE. We then extract the observed flux density in each of the $28$ \spirefts\ pointing sets for jiggle position $0$. The flux density from CIB sources is taken as the ensemble median over the set of $5000$ images in each pointing. These spectra are then averaged per detector, and the uncertainties of the averaged pointings are added in quadrature. The CIB estimates for each detector are shown in Figure \ref{fig:CIB}. For each CIB model, a diagonal covariance matrix is formed with diagonal elements given by the uncertainty estimated above. After subtraction of the CIB model, the final noise covariance matrix for these observations is formed from the sum of the noise covariance matrix estimated from the jackknives as described in Section~\ref{sec:noise_estimate} and the posterior variance of the CIB model.

\vspace{10pt}
\section{SZe Modeling and Inference}
\label{sec:fits}

\begin{figure*}[t!]
    \centering

    \includegraphics[width=0.45\linewidth]{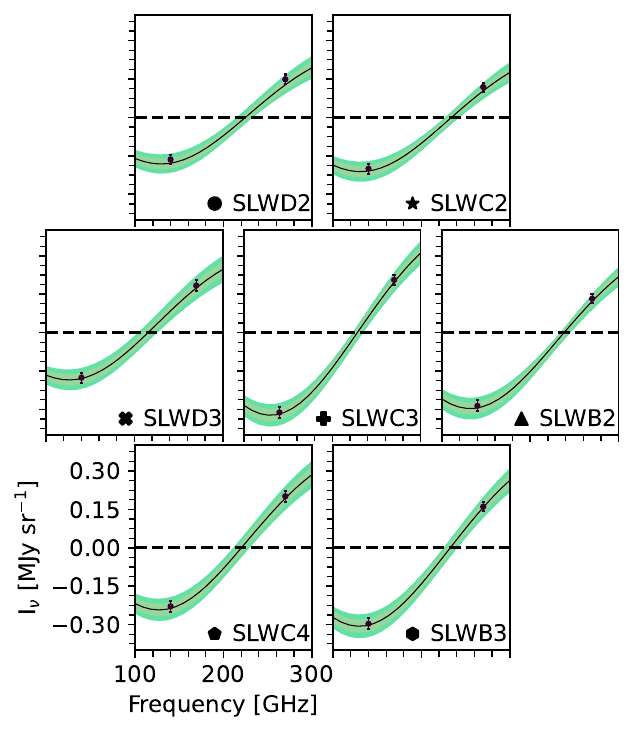} \quad
    \includegraphics[width=0.45\linewidth]{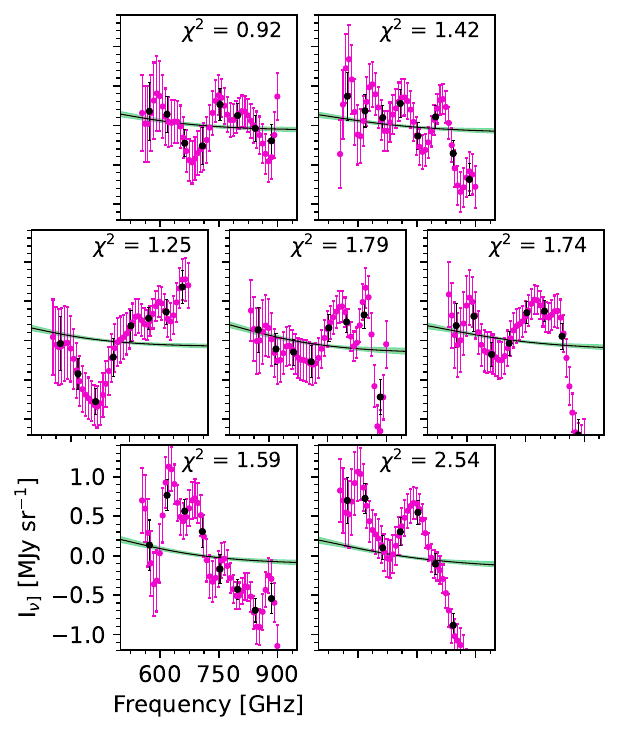}
    \caption{Shown here are best-fitting curves to the \SZe\ spectrum from the combined \spirefts\ (right) and Bolocam intensity measurements (left). The median fit is shown in black, with a $2\sigma$ confidence interval in green. The black data points on the right plots show the average of the \spirefts\ data, taken over a correlation length of six spectral channels. The uncertainty on these points is estimated by $\left(
    I^T CI\right)^{-1/2}$, where $I$ is the identity matrix, and $C$ is the covariance matrix sliced into $6\times6$ cutouts. The highest-frequency point is averaged over five spectral channels. The reduced $\chi^2$ values computed with the full covariance matrix are shown in each panel and in Table \ref{tab:best_fit_vals}. They imply much better fits than would be inferred directly from the pink data points due to the large correlations between adjacent bins. All panels share the same $x$ and $y$ axes which is depicted only for the lower-left subplots. }
    \label{fig:sze_spectral_fits}
\end{figure*}

\begin{table*}[htbp]
    \centering
    \begin{tabular}{c|c|c|c|c|c|c|c|c}
    \hline 
    Detector & SLWD2 & SLWC2 & SLWD3 & SLWC3 & SLWB2 & SLWC4 & SLWB3 & Units \\ \hline \hline
    MAP \TSZm & $\phantom{00}9.5\substack{+7.3\\-5.0}$ 
     & $\phantom{-}21.4\substack{+11.1\\-9.3}$ 
     &$\phantom{-}8.2\substack{+4.8\\-3.2}$ 
     & $\phantom{.}14.5\substack{+7.8\\-7.0}$ 
     & $\phantom{-}25.6\substack{+10.6\\-9.7}$ 
     & $\phantom{.}15.2\substack{+5.8\\-4.6}$ 
     & $\phantom{.}29.5\substack{+8.0\\-8.3}$ 
     & [keV] \\ 
    MAP $y$ & $17.0\substack{+2\\-2}$ 
     &$21.0\substack{+3\\-3}$ 
     & $19.0\substack{+2\\-2}$ 
     & $31.0\substack{+3\\-3}$ 
     & $29.0\substack{+4\\-3}$ 
     & $25.0\substack{+2\\-2}$ 
     & $33.0\substack{+3\\-3}$ 
     & [$\times 10^5$]\\ 
    MAP \ofts & $-0.06\substack{+0.02\\-0.02}$ 
     & $-0.1\substack{+0.03\\-0.03}$ 
     & $-0.08\substack{+0.01\\-0.02}$ 
     & $-0.17\substack{+0.03\\-0.04}$ 
     & $-0.14\substack{+0.04\\-0.03}$ 
     & $-0.11\substack{+0.02\\-0.02}$ 
     & $-0.18\substack{+0.03\\-0.03}$ 
     & [Mjy~sr$^{-1}$]\\ 
    MAP \vpec & $\phantom{0}140\substack{+180\\-160}$ 
     & $\phantom{.}840\substack{+280\\-280}$ 
     & $-400\substack{+110\\-110}$ 
     & $\phantom{.}340\substack{+140\\-140}$ 
     & $1470\substack{+170\\-170}$ 
     & $-540\substack{+90\\-90}$ 
     & $\phantom{.}240\substack{+110\\-110}$ 
     & [km~s$^{-1}$]\\ 

    $\chi^2$ & 0.920 & 1.421 & 1.245 & 1.785 & 1.740 & 1.589 & 2.537
 
    \end{tabular}
    \caption{Maximum a posteriori values for the parameters of our fit to each of the seven \spirefts\ detectors. Error bars indicate $68$\% credible regions. 
    }
    \label{tab:best_fit_vals}
\end{table*}

We model the total \SZe\ signal with
\begin{multline}
    \Delta I = \Delta I_{\KSZe}(\nu; y , \vpecm , T_{\rm rSZe})\\ + \Delta I_{\TSZe}(\nu; y, T_{\rm rSZe}) + \oftsm ,
\end{multline}
where \vpec\ is the peculiar velocity of the ICM, 
$y$ is the total Comptonization parameter, and \TSZm\ is the ICM temperature estimated by the \RSZe, which is assumed to be constant along the line of sight.
To marginalize over the absolute signal baseline of the \spirefts\ data, which is unconstrained, we include the term \ofts\ that describes an additive offset to the submillimeter data. We jointly fit the CIB-corrected \spirefts\ spectra shown in Figure \ref{fig:CIB} and the Bolocam $140$ and $270$ GHz data (described in \S \ref{sec:bolocam}) to this model to constrain the free parameters. The nuisance parameter \ofts\ is fit to the \spirefts\ data points only, and does not contain information of astrophysical importance. The data is fit with a MCMC sampler that evaluates $\Delta I_{\KSZe}$ and $\Delta I_{\TSZe}$ for each parameter combination using {\sc SZpack} \citep{SZpack1,SZpack2}. As part of this fit, we include uniform priors on \TSZm $ \in [0,60]$~keV and $y \in [0,1]$. The lower bounds are designed to ensure the MCMC does not sample negative values, which are not physically allowed. The upper bounds are designed to be minimally informative and do not noticeably impact the recovered posteriors. We also include an informative Gaussian prior on the value of \vpec\ based on the cluster-member redshifts derived in \S \ref{sec:velocity_data} and quantified in Table~\ref{tab:prior_values}, as well as absolute bounds of \vpec\ c$^{-1}$ $\in [-0.1,0.1]$. The way that HIPE processes the FTS data produces spectra that have an additive signal offset that is almost entirely determined by time-dependent properties of the instrument. While it is possible to place a prior on this offset based on expected total signal from the SZe, CIB, cirrus, and other relevant astrophysical sources, we found that these signal levels do not describe the observed offsets, so such a prior is not informative for our analysis. We thus treat the offset as a free parameter.

\begin{figure}[t]
    \centering
    \includegraphics[width=1\linewidth]{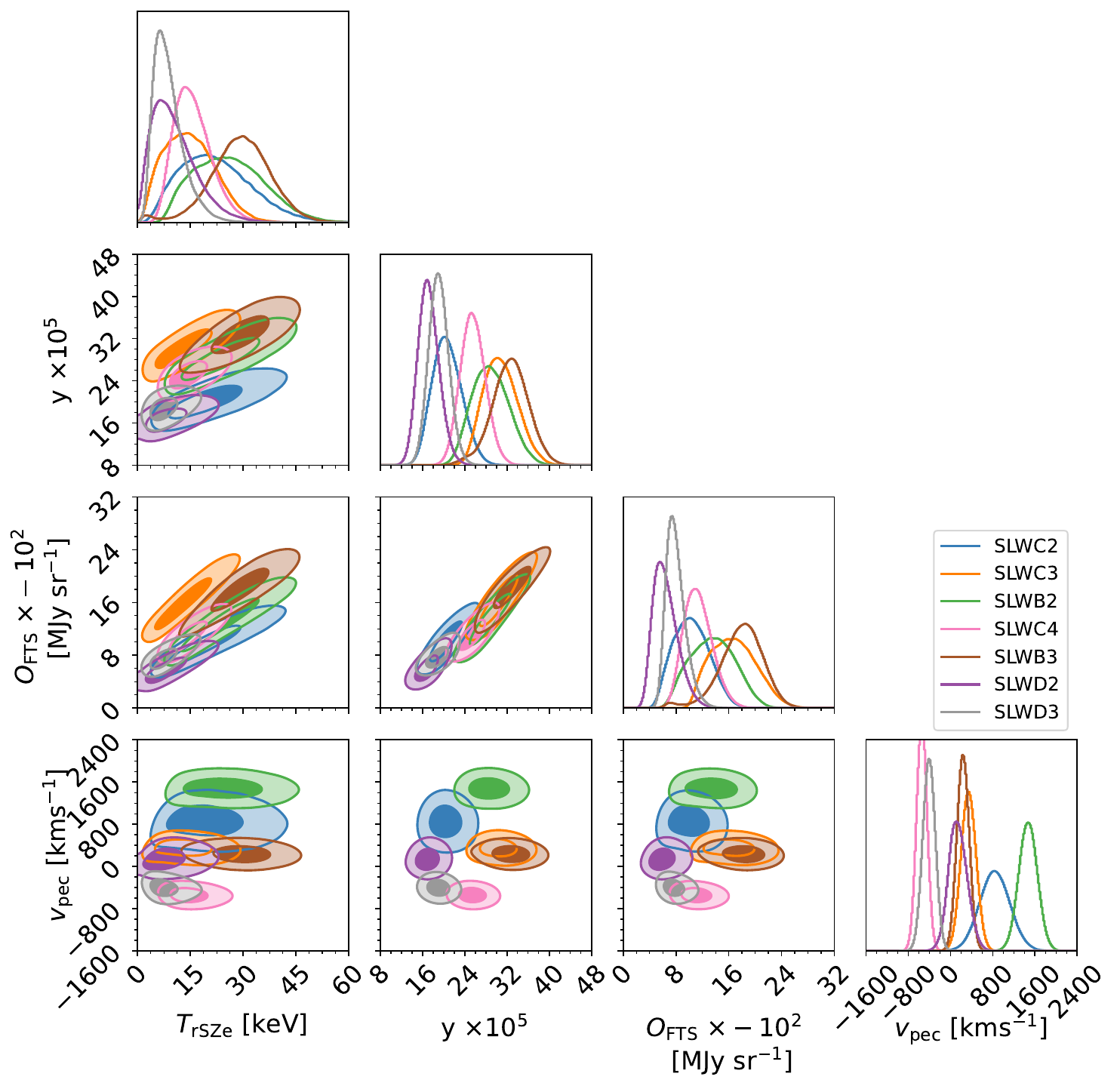}
    \caption{Posterior distributions for the fitted \SZe\ parameters for each of the seven \spirefts\ detector positions. Fitted spectra corresponding to these parameter values are shown in Figure \ref{fig:sze_spectral_fits}. The MAP values for each of the parameters being fit are shown in Table \ref{tab:best_fit_vals}, and can be compared to the fiducial values shown in Table \ref{tab:prior_values}.}
    \label{fig:corner_plots}
\end{figure}

The MCMC chains are initialized using random draws of the parameter values that sample the parameter space near the values of \TSZm, \vpec, and $y$ predicted from the fiducial values in Table \ref{tab:prior_values}. The initial draws for the nuisance parameter \ofts\ are $\in [-2.5,2.5]$ \mjysr. We run each of these chains for twenty thousand steps, and we cut the first three thousand steps as the burn-in samples. A Gelman-Rubin statistic \citep{gelman-rubin} is used to assess the convergence of the chains, where we find that $\hat{R} = 1.0$ and only has variation across all detector fits at the level of $10 \times 10^{-5}$. The posterior predictive from these fits is shown in Figure \ref{fig:sze_spectral_fits}, and the posterior constraints are shown in Figure \ref{fig:corner_plots} and in Table \ref{tab:best_fit_vals}.

If we assume that the ICM is isothermal, then the temperature measurements from the seven \spirefts\ detector positions can be combined to obtain an inverse variance weighted mean of \TSZm$=14.2\substack{+2.4\\-2.4}$~keV. However, the data strongly suggest that there are variations in ICM temperature over the cluster face. The fit quality of this isothermal model is relatively poor, as quantified by a reduced $\chi^2$ of $1.5$ and corresponding PTE of $0.06$. To account for such variations, we can extend the model to also include intrinsic scatter between the seven detector positions. We perform this fit with the \textit{LRGS} python package \citep{lrgs}, obtaining a mean temperature of \TSZm~$=15.1\substack{+3.8\\-3.3}$~keV with an intrinsic scatter of \sigrsze~$=5.4\substack{+5.1\\-3.4}$~keV.

 To assess potential biases in our fitting procedure, we created 100 mock data sets of the \RSZe\ data with realistic noise generated through a Cholesky decomposition. For each mock, the input value of \TSZm\ is drawn from a Gaussian distribution with $\mu=19.5$~keV and $\sigma=12$~keV. The other parameters are drawn from the posterior distributions shown in Figure \ref{fig:corner_plots} and in Table \ref{tab:best_fit_vals}. Each of these mocks is then to fit in the same manner as the \spirefts\ data, producing a set of $100$ best-fit temperatures. We resample these outputs with a jackknife method, wherein $10^4$ sets of seven measurements are aggregated via the \textit{LRGS} software package, following the same methodology as applied to the observed data. For each aggregate posterior, we extract the MAP value and produce a set of $10^4$ estimates. These values are then compared to the mean temperature of $19.5$~keV, and we find an on-average multiplicative bias of $1.016\substack{+0.003\\-0.003}$. We conclude that the bias in \TSZm\ from our fitting methodology is negligible compared to measurement uncertainties. Further, the recovered uncertainties on \TSZm\ in the mock estimates are comparable to those of the recovered data. 

\section{Discussion}
\label{sec:discussion}

Data from the \spirefts\ in combination with ground-based \SZe\ data from Bolocam have permitted a high-significance measurement of the ICM temperature in the quadruple merger \macs.  We find a cluster-average temperature of \TSZm$=15.1 \substack{+3.8\\ -3.3}$ keV with an intrinsic scatter over the cluster face corresponding to \sigrsze~$=5.4\substack{+5.1\\-3.4}$~keV. This work better establishes the \RSZe\ as a viable tool for probing ICM thermodynamics, particularly of the extremely hot gas expected in major mergers (and confirmed in \macs\ based on our analysis). In addition, while \macs\ is a moderate redshift source, our analysis also serves to advance the \RSZe\ as a potential path to probing the ICM in such systems at higher redshift. 

\subsection{Potential Differences between rSZE- and X-ray-Measured ICM Temperatures}
It is informative to compare the ICM temperature we obtain from the \RSZe\ against temperatures obtained from X-ray facilities. Differences between them may be expected from at least two sources. First, the differing line-of-sight weighting of the signals combined with nonisothermal distributions within the ICM can lead to discrepancies. Second, the presence of measurement systematics, in particular X-ray calibration uncertainties and band limitations, can also lead to different inferences about ICM temperature.

Regarding the first possibility, the \RSZe\ measurement is effectively pressure-weighted so it is biased toward any gas in post-shock regions that has extreme temperatures and pressures \citep[e.g.,][]{2019_Sze}. In contrast, the X-ray signal is biased toward high-density regions, which often correspond to low-temperature remnant cool cores in merging systems \citep[e.g.,][]{Rossetti2010}. In addition, differences are possible even in more quiescent systems, primarily as a result of cooling near the cluster center, resulting in an enhanced gas density within that region \citep[e.g.,][]{Hudson2010}. As a consequence, the X-ray signal will be biased toward these cooler central regions, which would result in a lower temperature compared to that derived from the \RSZe.

Regarding the second possibility, we start by noting that there is a well-established temperature calibration discrepancy between \xmm\ and \chandra, which was studied in particular detail by \citet{schellen}, who established that at least one satellite has a large systematic error in its calibration. In general, they find that \chandra\ obtains higher temperatures than \xmm, with a fractional difference of approximately $5$--$20$\% for typical ICM temperatures, with larger fractional differences at higher temperatures. This difference provides a characteristic scale for the possible systematic error that may exist in their temperature measurements. While more difficult to quantify, we also reiterate the band limitations of the X-ray facilities noted in Section~\ref{sec:intro}, which render them largely insensitive to photon energies $\gtrsim 8$~keV. Thus, both \xmm\ and \chandra\ have little to no sensitivity to the hottest regions that may be present in the ICM, which could produce lower X-ray-derived temperatures for such regions.

\subsection{Comparing rSZe and X-ray Measurements}
In this merging system, we expect both a very hot on-average ICM temperature as well as spatial variations in the temperature. To simultaneously assess such a temperature structure, we separately fit the \RSZe\ and X-ray data to a model that includes an overall mean temperature along with intrinsic scatter about this mean at the seven locations of the FTS pointings.

\begin{figure}[t]
    \centering
    \includegraphics[width=1\linewidth]{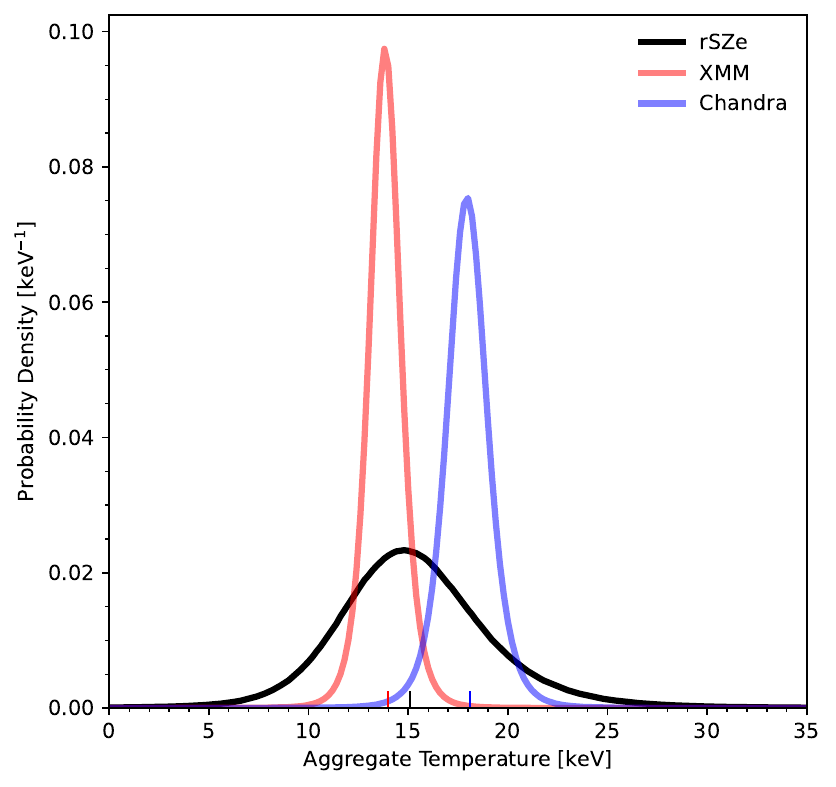}
         \caption{A comparison of the average temperatures obtained the \RSZe\ and X-ray analyses. The posterior distribution of \TSZm, accounting for detector-to-detector intrinsic scatter, is plotted in black, with \TSZm~$=15.1\substack{+3.8\\-3.3}$~keV. Applying the same procedure to the X-ray results yields \TXMM~$=13.9\substack{+0.9\\-0.9}$~keV and \TChan~$=18.0\substack{+1.1\\-1.1}$~keV.}
    \label{fig:aggregate_temperature}
\end{figure}

To compute a characteristic X-ray temperature in a manner that best facilitates comparison to our \RSZe-derived measurement, we determine the average temperature from the X-ray maps shown on the bottom row of Figure~\ref{fig:all-of-the-data} at the same locations sampled by the \spirefts\ detectors, correcting for the PSF sampling according to the procedure described in Section~\ref{sec:data_extract}. This yields a cluster-average temperature of \TChan$= 18.0\substack{+1.1\\-1.1}$~keV from \chandra\ and \TXMM$=13.9\substack{+0.9\\-0.9}$~keV from \xmm, shown as Fig.~\ref{fig:aggregate_temperature}. While these values appear to be in mild tension, we note that the difference is almost exactly what is expected from the empirical relation derived by \citet{schellen}, and thus we attribute this tension solely to calibration uncertainties. The corresponding intrinsic scatters over the cluster face are \sigchan~$=2.8\substack{+1.3\\-0.8}$~keV, and \sigxmm~$=2.2\substack{+1.0\\-0.6}$~keV.

Combining the results of all three analyses, we find consistent average temperatures in the range of 15--20~keV for \macs. We firmly and independently establish that this major merger contains a significant amount of shock-heated gas from our \RSZe\ analysis. In addition, other than the known calibration difference between \xmm\ and \chandra, we find no evidence for differences in average temperature e.g.,~due to the specific distribution of gas density and temperature in \macs\ or to potential biases from the band limitations of the X-ray facilities. While it may be tempting to utilize these measurements to better establish the temperature calibration of the X-ray satellites, we assert that such an inference is not justified due to the astrophysical and instrumental effects noted above, which are particularly relevant for an extremely hot merging system like \macs\ even if they are not found to be statistically significant in our specific analysis. Further caution against such a comparison is warranted due to the lack of a homogeneous analysis of the \chandra\ and \xmm\ data. Finally, we obtain a nonzero intrinsic scatter at modest statistical significance and with a characteristic value of 2--5~keV from all three probes. This suggests that the ICM in \macs\ is not isothermal, and the variations in temperature over the projected face of the cluster are likely attributable to the ongoing multiple merger.

While we caution against utilizing this present analysis as a tool to calibrate X-ray temperature measurements, we note that the \RSZe\ does hold significant promise for obtaining such a calibration assessment. For instance, \citet{Biffi2014} find an on-average negligible bias between ICM temperatures obtained from different probes when restricting to morphologically regular clusters. Thus, an analysis based on a large sample of regular clusters would largely eliminate the potential astrophysical systematics related to ICM substructures noted above. In addition, the temperature of the ICM in such clusters should fall within the energy bands of the X-ray satellites.

\subsection{Prospects for Future Spectroscopic Measurements of rSZe}

Looking to future spectrophotometric measurements of the \SZe, this work has yielded some interesting insights.  First, while FTS systems offer high spectral resolving power and potentially large mapping speeds, they are also very sensitive to mechanical and thermal changes that impact the optical path length while acquiring data.  The \spirefts, in particular, was not designed to be sensitive to slowly varying continuum signals, and many of the low-frequency systematic effects in the data appear to be at least in part due to changes in the thermal properties of the instrument or the characteristics of the mechanical scanning mechanism.  Optical systematic errors are also problematic in these measurements, with sensitivity to the precise temperature and configuration of the cold optics resulting in significant data loss.  The inclusion of null data in the Dark Sky field to characterize these effects was crucial to extracting a scientific result from these observations.  Secondly, as is made clear in Fig.~\ref{fig:sze_sensitivity}, measurements of the \RSZe\ should be optimized for bandwidths of $\sim 50 \,$GHz ($R \sim 15$) at submillimeter wavelengths.  This would optimize sensitivity to the broad continuum signal produced by the \SZe, while allowing sensitivity to the spectral features in the \RSZe\ deviation that provide contrast to statistically constrain the temperature of the ICM.  Finally, this case study highlights the need for accurate information about the submillimeter galaxies in the cluster field.  The \RSZe\ effect would have been buried beneath the CIB sources if we did not have high-sensitivity broadband information to subtract their contribution to the surface brightness.  

\subsection{Conclusion}

We were able to measure the \RSZe\ signal from \macs\ at seven unique locations with the SPIRE Fourier transform spectrometer. These measurements were then aggregated by fitting a characteristic mean and intrinsic scatter, where we find an \RSZe\ inferred average temperature of \TSZm~$=15.1\substack{+3.8\\-3.3}$~keV which is consistent with X-ray inferred temperatures from both \chandra\ and \xmm. While this very deep space-based FTS measurement is likely to remain unique for the foreseeable future, we have demonstrated that interesting new measurements of the ICM using the \RSZe\ effect are well within the range of a sensitive next-generation instrument that has carefully controlled systematics and modest spectral capability. Measurements of this nature offer a complementary probe of both high-temperature and high-redshift ICM gas, which would allow for detailed study of the evolution of cluster thermodynamics that would otherwise be difficult or nonpermissible to understand utilizing other methods.

\section*{Acknowledgements}
The authors sincerely thank I.~Valtchanov for his assistance designing, executing, and understanding the \spirefts\ observations as our ``Friend of the Observer'' many years ago.

This work has been supported in part by
the National Aeronautics and Space Administration under grants NAS7-03001-1467865, 80NSSC19K1018, and 80NSSC24K1024 and by the National Science Foundation under grants DGE-1745301 and NSF/AST-2206082.  L. Lovisari acknowledges support from INAF grant 1.05.24.05.15.  The authors acknowledge Research Computing at the Rochester Institute of Technology for providing computational resources and support that have contributed to the research results reported in this publication.

The scientific results reported in this article are based in part on observations made with Herschel, a European Space Agency Cornerstone Mission with significant participation by NASA.  HIPE is a joint development by the Herschel Science Ground Segment Consortium, consisting of ESA, the NASA Herschel Science Center, and the HIFI, PACS and SPIRE consortia.  This research has made use of the NASA/IPAC Infrared Science Archive, which is funded by the National Aeronautics and Space Administration and operated by the California Institute of Technology.

Some of the data presented herein were obtained at the W. M. Keck Observatory, which is operated as a scientific partnership among the California Institute of Technology, the University of California, and NASA. The Observatory was made possible by the generous financial support of the W. M. Keck Foundation. The authors wish to recognize and acknowledge the very significant cultural role and reverence that the summit of Maunakea has always had within the indigenous Hawaiian community. We are most fortunate to have the opportunity to conduct observations from this mountain.

The scientific results reported in this article are based  in part on work at the Caltech Submillimeter Observatory, which is operated by the California Institute of Technology.

The scientific results reported in this article are based  in part on observations obtained with \xmm, an ESA science mission with instruments and contributions directly funded by ESA Member States and NASA.

This paper employs a dataset, obtained by the \chandra\ X-ray Observatory, contained in the \chandra\ Data Collection ~\dataset[DOI: 10.25574/1655]{https://doi.org/10.25574/cdc.548}.

\facility{Herschel, CSO, XMM, CXO, Keck I, and Keck II}

\software{astropy, \citep{astropy_2013,astropy_2018, astropy_2022}, SZpack \citep{SZpack1, SZpack2}, emcee \citep{emcee}, PCAT \citep{PCAT_2023}, scipy \citep{scipy}, Xspec \citep{xspec}, Numpy \citep{numpy}, contbin \citep{contbin},  HIPE \citep{hipe_mandatory}, matplotlib \citep{matplotlib}, LMC}

\bibliographystyle{aasjournal}
\bibliography{sample631}{}

\end{document}